\documentclass[11pt]{article}
\usepackage[dvips]{graphicx}
\usepackage{latexsym}
\usepackage{euscript}

\newcommand{\MSbar}{\overline{\mbox{MS}}}
\newcommand{\RI}{\mbox{RI}}
\newcommand{\ri}{\mbox{\scriptsize RI}}

\newcommand{\Tr}{\mbox{Tr}\;} 

\renewcommand{\>}{\rangle}
\newcommand{\bc}{\begin{center}}
\newcommand{\ec}{\end{center}}
\newcommand{\be}{\begin{equation}}
\newcommand{\ee}{\end{equation}}
\newcommand{\bea}{\begin{eqnarray}}
\newcommand{\eea}{\end{eqnarray}}
\newcommand{\ba}{\begin{eqnarray}}
\newcommand{\ea}{\end{eqnarray}}
\newcommand{\brr}{\begin{array}}
\newcommand{\err}{\end{array}}

\newcommand{\nn}{\nonumber}
\newcommand{\id}{\hbox{1$\!\!$1}}

\newcommand{\simge}{\ \lower-
1.2pt\vbox{\hbox{\rlap{$>$}\lower5pt
\vbox{\hbox{$\sim$}}}}\ }

\begin{document}
\pagestyle{empty} 
\vspace{-0.6in}
\begin{flushright}
BUHEP-00-17\\
MIT-CTP-3029\\
November 2000
\end{flushright}
\vskip 1.5in

\centerline{\large {\bf{Analysis of the $\Delta I =1/2$ Rule 
and $\epsilon'/\epsilon$}}} 
\centerline{\large {\bf{ with Overlap Fermions}}}
\vskip 0.6cm
\centerline{\bf Stefano Capitani$^{1}$, Leonardo Giusti$^{2}$}
\vskip 0.5cm
\centerline{$^1$ Center for Theoretical Physics, Laboratory for Nuclear Science}
\centerline{Massachusetts Institute of Technology}
\centerline{77 Massachusetts Avenue, Cambridge MA 02139, USA}
\centerline{e-mail: stefano@mitlns.mit.edu}
\vskip 0.2cm
\centerline{$^2$ Department of Physics - Boston University} 
\centerline{590 Commonwealth Avenue, Boston MA 02215, USA}
\centerline{e-mail: lgiusti@bu.edu}
\vskip 1.0in
\begin{abstract}
We study the renormalization of the $\Delta S =1$ effective 
weak Hamiltonian with overlap 
fermions. The mixing coefficients among 
dimension-six operators are computed at one loop in perturbation theory. 
As a consequence of the chiral symmetry at finite 
lattice spacing and of the GIM mechanism, which is 
quadratic in the masses, the $K\rightarrow \pi\pi$ and 
$K\rightarrow \pi$ matrix elements relevant for the $\Delta I =1/2$ rule 
can be computed without any power subtractions. The analogous amplitudes 
for $\epsilon'/\epsilon$ require  
one divergent subtraction only, which can be performed non-perturbatively
using $K\rightarrow 0$ matrix elements.
\end{abstract}
\begin{flushleft}
\ \ \ \ \ \ \ \ \ \ \ \ \ 11.15.Ha, 12.38.Gc, 14.40.Aq
\end{flushleft}
\vfill
\pagestyle{empty}\clearpage
\setcounter{page}{1}
\pagestyle{plain}
\newpage 
\pagestyle{plain} \setcounter{page}{1}

\newpage

\section{Introduction}
The non-leptonic $\Delta S =1$ weak transitions are far from being 
understood. It is well known that the $\Delta I=1/2$ amplitudes\footnote{
In this paper we will study only amplitudes for $K\rightarrow \pi\pi$ decays.}
are enhanced (the so-called $\Delta I=1/2$ rule) with respect 
to the corresponding $\Delta I =3/2$ ones 
(by roughly a factor $20$ in $K\rightarrow \pi\pi$ decays); the latest 
measurements of $\epsilon'/\epsilon$~\cite{epp/ep_exp}
confirm the rather large value found by NA31
\cite{NA31_old}, and the up-to-date world average is roughly $2 \cdot
10^{-3}$. Given the present limited knowledge of the long-distance QCD effects in 
non-leptonic $\Delta S =1$ amplitudes, these experimental 
results cannot be predicted from first principles and 
well-controlled approximations. 

The short-distance QCD corrections to the   
$\Delta S =1$ effective Hamiltonian can be reliably computed in 
perturbation theory\footnote{In the last 
few years there has been much progress in constructing a non-perturbative 
formulation of chiral gauge theories on the lattice 
\cite{luscher2,golterman}.} \cite{deltaI=1/2}.
In the Standard Model (SM) the Wilson 
coefficients are known up to the Next-to-Leading Order (NLO) 
\cite{buras,reina}. They account for a factor two of 
the enhancement in the $\Delta I =1/2$ rule and  
cannot reproduce the experimental value of $\epsilon'/\epsilon$, when 
$\mbox{Im}\lambda_t\equiv V_{td}V^*_{ts}$ is determined from the unitarity triangle 
and the Vacuum Saturation Approximation (VSA) estimate of 
$\langle \pi\pi |O_6 |K\rangle$ (see below) is used \cite{silvester}-\cite{pp}. 
The SM could explain  both effects if penguin contractions,
neglected in the VSA, give contributions to
the relevant matrix elements definitely larger than their 
factorized values \cite{ep/e}. Up to now no reliable first-principle 
computations of the $\Delta I=1/2$ matrix elements exist. 

Lattice QCD is the only known method which allows one to compute non-perturbative 
QCD contributions to physical amplitudes from first principles. 
In the most popular lattice regularizations, i.e.~Wilson and staggered
fermions, the main theoretical aspects of the renormalization of 
composite operators are fully understood~\cite{boc}-\cite{tassos}. 
The results of quenched calculations of the matrix elements relevant for the
$K^0$--$\bar K^0$ and $B^0$--$\bar B^0$ mixings  have
reached a high level of accuracy and are commonly used in
phenomenological analyses of the unitarity 
triangle~\cite{sharpe}-\cite{NoiDELTAS=2}.
These lattice results have been crucial in constraining the 
parameters of the CKM matrix in the Standard Model (SM) 
\cite{CKM,ep/e} and beyond \cite{CKM_beyond}. 
Techniques to compute $\Delta I =1/2$ amplitudes 
for the $\Delta S=1$ Hamiltonian 
have been developed for both Wilson and staggered
fermions \cite{bs,old_lat}, but these methods have not yet produced useful
results\footnote{Attempts with domain-wall fermions are in progress \cite{blum}.}.
There are two major 
difficulties:
\begin{itemize}
\item In Euclidean space there is apparently no simple relation between 
      the two-~(many-)body transition matrix elements and the corresponding
      correlation functions at large time separation \cite{maiani_testa}. 
      This problem is common to all lattice discretizations and there are 
      proposals to solve it either by using finite volume techniques
      \cite{LaurentLuscher} or studying the correlation functions
      in the large volume limit \cite{sacmarti,testa_new};
\item The operators we are interested in 
      mix with lower-dimensional operators with coefficients which can
      diverge as inverse powers of the lattice spacing. These mixings 
      can be more severe when chiral symmetry is broken by the 
      regularization, like for Wilson fermions. 
\end{itemize} 
In the last few years it has been understood \cite{neub1}-\cite{luscher} 
that chiral and flavor symmetries can be preserved simultaneously on the
lattice, without fermion doubling, if the fermionic operator $D$ satisfies the 
Ginsparg-Wilson Relation (GWR) \cite{GW}. This implies an exact
invariance of the fermionic action which can be interpreted as a
lattice realization of the standard chiral symmetry 
at finite cutoff \cite{luscher}. However the GWR itself does not
guarantee locality, 
the absence of doubler modes  and the correct classical continuum limit. 

Neuberger, through the overlap formalism \cite{neub0}, found a solution \cite{neub1} 
of the GWR which satisfies all the above requirements and is local\footnote{
The overlap-Dirac operator is not ultra-local. The Neuberger 
kernel satisfies a more general definition of locality, i.e. it 
is exponentially suppressed at large distances with a decay rate 
proportional to $1/a$.} \cite{pilar}.
The complicated form of Neuberger's operator  
renders its numerical implementation quite demanding
for the present generation of computers. However, some progress has been achieved 
\cite{simulazioni} and Monte Carlo simulations are already 
feasible, at least in the quenched approximation.

In this paper we study the renormalization of the 
$\Delta S =1$ effective weak Hamiltonian in the overlap 
regularization. Neuberger's action 
implies many theoretical advantages:
\begin{itemize}
\item The quark masses renormalize only multiplicatively, the  
mixings among operators with different chirality are forbidden
\cite{hasenfratz2,capgiu} and their parity-violating and 
parity-conserving components renormalize in the same way \cite{capgiu};
\item The GIM mechanism is as powerful as in the continuum to suppress 
      mixing with lower-dimensional operators;  
\item The action and therefore the spectrum of the theory are free from
  $O(a)$ discretization errors. The $O(a)$ improvement of the local 
  fermionic operators is greatly simplified~\cite{qcdsf}.
\end{itemize}
As a consequence the $K\rightarrow \pi\pi$ and, if chiral
perturbation theory ($\chi$PT) is used,  $K\rightarrow \pi$ matrix elements 
for the $\Delta I=1/2$ rule can be computed without any power 
subtractions. The analogous matrix elements 
for $\epsilon^\prime/\epsilon$ require only
one divergent subtraction which can be performed non-perturbatively
using on-shell $K\rightarrow 0$ matrix elements. On the contrary,
when the regularization breaks chiral symmetry the $K\rightarrow\pi$ matrix 
elements for the $\Delta I =1/2$ rule 
and for $\epsilon'/\epsilon$ require one and two power-divergent 
subtractions respectively.

We also compute the mixing coefficients among dimension-six operators
at one loop in perturbation theory\footnote{These renormalization 
constants for domain-wall fermions in the limit of infinite fifth dimension 
have been computed in \cite{aoki}.}. A further determination of these 
renormalization constants can be obtained using  numerical 
non-perturbative methods such as those in Refs.~\cite{NP,luscher_np}. 
However a perturbative analysis is useful for analytically studying
the mixing pattern of the operators, 
often gives accurate estimates of the renormalization constants  
and furnishes a consistency check of the non-perturbative methods. 

The paper is organized as follows: in section \ref{sec:continuum} we 
define the $\Delta S =1$ effective Hamiltonian in the continuum;
in sections \ref{sec:definitions} and \ref{sec:DS1_renormalization}  
the overlap fermion action and the lattice 
$\Delta S =1$ operators are introduced, and
in section  \ref{sec:penguins} we compute at one loop the penguin diagrams
of these operators. 
In sections \ref{sec:DI=1/2} and \ref{sec:ep/e} we 
describe the renormalization pattern of the relevant  operators 
for the $\Delta I =1/2$ rule and $\epsilon'/\epsilon$ and we report 
the main results of this paper; in section \ref{sec:conclusions} we 
state our conclusions.  

\section{The $\Delta S =1$ effective Hamiltonian in the continuum}\label{sec:continuum}
The $\Delta S=1$ effective Hamiltonian above the charm threshold
is given by
\bea
{\cal H}_{eff}^{\Delta S =1}&=&\lambda_u \frac {G_F} {\sqrt{2}}
\Bigl[ (1 - \tau ) \Bigl( C_1(\mu)\left( \widehat Q_1(\mu) - \widehat Q_1^c(\mu) \right) +
C_2(\mu)\left( \widehat Q_2(\mu) - \widehat Q_2^c(\mu) \right)  \Bigr)\nn\\
&+&\tau \sum_{i=1}^{10} C_i(\mu) \widehat Q_i(\mu) \Bigr]\; ,
\protect\label{eq:eh}
\eea
where $G_{ F}$ is the Fermi coupling constant,
$\tau=-\lambda_t/\lambda_u$,
$\lambda_u = V_{ud} V^*_{us}$, $\lambda_t= V_{td} V^*_{ts}$
and $\mu$ is the renormalization scale of the composite operators.
The Wilson coefficients $C_i(\mu)$ at the NLO in QCD and QED 
have been given in \cite{buras,reina}. The
operator basis is
\bea
Q_{1}&=&({\bar s}^{\alpha} \gamma^\mu_{L} u^{\beta} )
        ({\bar u}^{\beta}  \gamma^\mu_{L} d^{\alpha})
\nonumber\\
Q_{2}&=&({\bar s} \gamma^\mu_{L} u)
        ({\bar u}  \gamma^\mu_{L} d)
\nonumber \\
Q_{3,5} &=& ({\bar s}\gamma^\mu_{L} d)
    \sum_{q}({\bar q}\gamma^\mu_{L,R}q)
\nonumber \\
Q_{4,6} &=& ({\bar s}^{\alpha}\gamma^\mu_{L} d^{\beta})
    \sum_{q}({\bar q}^{\beta}\gamma^\mu_{L,R} q^{\alpha})
\nonumber \\
Q_{7,9} &=& \frac{3}{2}({\bar s}\gamma^\mu_{L} d)
    \sum_{q}e_{ q}({\bar q}\gamma^\mu_{R,L} q)
\nonumber \\
Q_{8,10} &=& \frac{3}{2}({\bar s}^{\alpha}\gamma^\mu_{L}d^{\beta})
    \sum_{q}e_{ q}({\bar q}^{\beta}\gamma^\mu_{R,L} q^{\alpha})
\nonumber \\
Q^c_{1}&=&({\bar s}^{\alpha}\gamma^\mu_{L}c^{\beta} )
          ({\bar c}^{\beta}\gamma^\mu_{L}d^{\alpha})
\nonumber \\
Q^c_{2}&=&({\bar s} \gamma^\mu_{L} c)
          ({\bar c} \gamma^\mu_{L}d )\; ,
\protect\label{epsilonprime_basis}
\eea
where 
$\gamma^\mu_{L,R} = \gamma^\mu(1\mp \gamma_5)$, $e_q$ are the quark electric charges 
and $\alpha,\beta$ denote color indices (which are omitted for the
color-singlet operators). 
Below the bottom
threshold, $q$ runs over the active flavors $u,d,s,c$.  
Other operators would appear in Eq.~(\ref{eq:eh}), but 
in the Standard Model
they are numerically irrelevant  and 
they will not be considered in this paper\footnote{In 
some supersymmetric extensions of the SM, large contributions to 
$\epsilon'/\epsilon$ can be induced by the magnetic and chromomagnetic 
operators \cite{masmur}.}. 

The electro-penguin operators $Q_{i}$ ($i=7,\dots,10$)
give significant contributions to 
$\epsilon'/\epsilon$ only through their 
$\Delta I =3/2$ components. In the limit 
$m_u=m_d$, $Q^{3/2}_{i}$ do not mix with lower-dimensional 
operators\footnote{Estimates for the 
matrix elements of these operators in the quenched approximation 
have been obtained on the lattice in the standard 
regularizations \cite{sharpe}-\cite{NoiDELTAS=2}.} 
and their renormalization constants in the overlap regularization 
can be obtained from \cite{capgiu}. 
Therefore in the following we will focus only on the computations
of the matrix elements of 
$Q_1,\dots,Q_6$ (and $Q^c_1,Q^c_2$).   

Under flavor $SU(3)_L\times SU(3)_R$ transformations, the combination 
\be\label{ref:ventisette}
Q_{27} = 2 Q_2 + 3 Q_1 - Q_3 + Q^c_1
\ee
transforms as a $(27_L,1_R)$ multiplet and induces 
both $\Delta I=1/2$ and $\Delta I =3/2$ transitions.
The operators 
\be
Q_{-} =Q_1 - Q_2
\ee
and $Q_i$ ($i=3,4,5,6$) transform as $(8_L,1_R)$ operators and
induce pure $\Delta I=1/2$ transitions.

The octet operators are linearly dependent; by 
performing Fierz transformations it is easy to show that
\be\label{eq:dipendenza}
Q_4 = Q_3 + Q_2 - Q_1 + Q_2^c - Q_1^c\; .
\ee
For $\mu<m_c$ the operator basis is given by
$Q_1-Q_{10}$ and the sums over $q$ in the penguin 
operators run over $u,d,s$ only. The relation (\ref{eq:dipendenza})
becomes
\be\label{eq:dipendenza2}
Q_4 = Q_3 + Q_2 - Q_1 
\ee
and in Eq.~(\ref{ref:ventisette}) $Q^c_1$ has to be dropped.

\section{Basic definitions for overlap fermions}\label{sec:definitions} 
The QCD lattice action we consider for massive
fermions is  
\be   
S_L = \frac{6}{g_0^2}\sum_{P} \bigg[ 1 - \frac{1}{6}   
\Tr \Big[ U_P + U_P^{\dagger} \Big] \bigg]  + 
\bar{q}
\left[\Big(1-\frac{1}{2\rho}a m_q\Big)D_{N} + m_q\right] \, q,
\label{eq:sg}   
\ee   
where, in standard notation, $U_P$ is the Wilson plaquette, 
$g_0$ is the bare coupling constant and $m_q$ is the bare 
fermion mass. $D_{N}$ is the Neuberger-Dirac operator defined as
\ba\label{eq:opneub}
D_N &=& \frac{\rho}{a}\left( 1 + X\frac{1}{\sqrt{X^\dagger X}}\right)\nonumber\\
X &=& D_W -\frac{1}{a}\rho\; ,
\ea
where 
\be\label{eq:WDO}
D_W = \frac{1}{2} \gamma_\mu (\nabla_\mu + \nabla^*_\mu)
-\frac{r}{2} a \nabla^*_\mu\nabla_\mu
\ee
is the Wilson-Dirac operator\footnote{All perturbative
computations reported in this paper are performed with $r=1$.}, $0<r\leq 1$ and 
$0<\rho< 2r$. $\nabla_\mu$ and
$\nabla^*_\mu$ are the forward and backward lattice
derivatives, i.e.
\ba
\nabla_\mu q(x) & = & \frac{1}{a}\Big[U_\mu(x) q(x + a \hat \mu) -
q(x)\Big]\nonumber\\
\nabla_\mu^* q(x) & = & 
\frac{1}{a}\Big[ q(x) - U^{\dagger}_\mu(x-a \hat \mu) 
q(x-a\hat\mu) \Big]\; ,\label{eq:derivative}
\ea
where $U_\mu(x)$ are the lattice 
gauge links. 

The overlap-Dirac operator satisfies the Ginsparg-Wilson relation
\be\label{eq:GW}
\gamma_5 D_N + D_N \gamma_5 = \frac{a}{\rho} D_N \gamma_5 D_N  
\ee
which, in the massless limit, implies a continuous 
symmetry of the action which may be interpreted as a lattice form 
of the chiral invariance \cite{luscher}:
\be\label{eq:luscher_new}
\delta q = \hat \gamma_5 q\; , \qquad  
\delta \bar q = \bar q \gamma_5  \; ,
\ee
where $\hat \gamma_5$ is defined as
\be
\hat \gamma_5 = \gamma_5\Big(1-\frac{a}{\rho}D_N\Big)
\ee
and satisfies
\be
\hat \gamma_5^\dagger = \hat\gamma_5\; , \qquad \hat \gamma_5^2 = 1\; .
\ee
If we define the chiral projectors as 
\be
\hat P_{\pm} = \frac{1}{2}(1\pm \hat \gamma_5)\; , \qquad 
P_{\pm} = \frac{1}{2}(1\pm \gamma_5)\; ,
\ee
the fermionic fields can be decomposed into left- and right-handed 
components \cite{luscher2,narayanan,luscher3}  
\be
q_{R,L} = \hat P_{\pm} q \qquad \bar q_{R,L} = \bar q P_{\mp} 
\ee
which transform under lattice chiral rotations 
in the same way as the corresponding fields in the continuum.
The massive fermion matrix in Eq.~(\ref{eq:sg}) can be written as
\be
D_N(m) = P_+ D_N \hat P_- + P_- D_N \hat P_+ + m (P_- \hat P_- + P_+ \hat
P_+) \; .
\ee 
The composite fermionic operators can be defined
to have the same combination of left and right components as in the 
continuum. This guarantees that under lattice 
chiral rotations
they transform as the corresponding operators in the continuum.
As an example, the bilinear operators can be defined as  
\be\label{eq:bilinears}
{\cal O}_\Gamma = \bar q_1 \bar\Gamma q_2\; , 
\ee
where $\bar{\Gamma}=\Gamma(1-\frac{a}{2\rho} D_N)$. The on-shell matrix 
elements of these operators are $O(a)$ improved \cite{qcdsf}.
The Feynman rules corresponding to the action in Eq.~(\ref{eq:sg})
are given in \cite{capgiu} and are in agreement with those 
used in \cite{japan}-\cite{stefano_neub}. Reference~\cite{capgiu}
also gives in detail all conventions and symbols used in this paper.

\section{$\Delta S=1$ four-fermion operators on the lattice}
\label{sec:DS1_renormalization}
The four-flavor QCD lattice action in the overlap regularization reads
\be
S = \frac{6}{g_0^2}\sum_{P} \bigg[ 1 - \frac{1}{6}   
\Tr \Big[ U_P + U_P^{\dagger} \Big] \bigg]  + 
\sum_q \bar{q} \left[\Big(1-\frac{1}{2\rho}a m_q\Big)D_{N} + m_q\right] \, q,
\label{eq:sg_QCD}      
\ee
where $q=u,d,s,c$. In the limit $m_u=m_d=m_s=0$, 
this action is 
invariant under the $SU(3)_L\times SU(3)_R$ group and 
the corresponding transformations are defined
as 
\ba\label{eq:non_sing_chiral}
\psi_L \rightarrow U \psi_L\; , & \qquad & 
\bar \psi_L \rightarrow \bar \psi_L U^{\dagger}\nonumber\\
\psi_R \rightarrow V \psi_R\; , & \qquad & 
\bar \psi_R \rightarrow \bar \psi_R V^{\dagger}\; ,
\ea
where $\psi_{L,R}=(u_{L,R},d_{L,R},s_{L,R})$ is a column vector in flavor space
and 
\ba
U^{\dagger}U & = & V^{\dagger}V =\id\nonumber\\
\mbox{det} U & = & \mbox{det} V = 1\; .
\ea  
When $u,d,s$ acquire a mass, the action is invariant if 
the light quark masses are transformed as
\be
M =   
\left(
\begin{array}{ccc} 
m_u &     &     \\
    & m_d &     \\
    &     & m_s \\
\end{array}\right)
\rightarrow U M V^{\dagger}\; .
\ee
A convenient definition for the lattice  
four-fermion operators of the $\Delta S =1$ effective Hamiltonian is
\ba\label{eq:basis_lattice}
{\cal Q}_1 & = & ({\bar s}^{\alpha}\bar \gamma^\mu_L u^{\beta})
({\bar u}^{\beta}\bar \gamma^\mu_L d^{\alpha})\nonumber\\
{\cal Q}_2 & = & ({\bar s}\bar \gamma^\mu_L u)
({\bar u} \bar \gamma^\mu_L d) \nonumber\\
{\cal Q}_{3,5} & = & ({\bar s}\bar \gamma^\mu_L d)
\sum_q ({\bar q}\bar \gamma^\mu_{L,R} q)\\
{\cal Q}_{4,6} & = & ({\bar s}^{\alpha}\bar \gamma^\mu_L d^{\beta})
\sum_q ({\bar q}^{\beta}\bar \gamma^\mu_{L,R} q^{\alpha})\nonumber\\
{\cal Q}^c_1 & = & ({\bar s}^{\alpha}\bar \gamma^\mu_L c^{\beta}) 
({\bar c}^{\beta}\bar \gamma^\mu_L d^{\alpha})\nonumber\\
{\cal Q}^c_2 & = & ({\bar s}\bar \gamma^\mu_L c)
({\bar c} \bar \gamma^\mu_L d) \; ,\nonumber
\ea
where $\bar \gamma^\mu_{L,R} = \gamma^\mu_{L,R}(1-\frac{a}{2\rho} D_N)$.
Under the rotations in Eq.~(\ref{eq:non_sing_chiral}), 
these operators transform as the analogous operators in the continuum
and their on-shell matrix elements are $O(a)$ improved. 

Under renormalization the operators in 
Eqs.~(\ref{eq:basis_lattice}) can mix
with equal- or lower-dimensional operators which have the same
transformation properties under $SU(3)_L\times SU(3)_R$ or, due to the 
explicit soft chiral symmetry-breaking term, with multiplets of different 
chirality properly multiplied by light quark masses. As in the continuum, 
${\cal Q}_{27}$ is multiplicatively renormalizable
while the 
$(8_L,1_R)$ operators can mix among themselves as well as with 
two lower-dimensional operators \cite{old_lat,bdspw}
(besides operators which vanish by the equations of motion)
\ba
{\cal Q}_\sigma & = & g_0\left[(m_s+m_d)\bar s \bar\sigma_{\mu\nu} F_{\mu\nu} d
+ (m_s-m_d) \bar s \bar \sigma_{\mu\nu} \widetilde F_{\mu\nu} d\right]\label{eq:Qsigma}\\
{\cal Q}_m & = & (m_s+m_d)\bar s \bar{\id} d + (m_s-m_d) 
\bar s \bar \gamma_5 d\; ,\label{eq:Qm}
\ea
where $F_{\mu\nu}$ and 
$\widetilde F_{\mu\nu}$ are the non-abelian gluon field tensor and its dual, 
$\bar\sigma_{\mu\nu} = \sigma_{\mu\nu}(1-\frac{a}{2\rho} D_N)$, 
$\bar \gamma_5 = \gamma_5 (1-\frac{a}{2\rho} D_N)$ and 
$\bar{\id} = (1-\frac{a}{2\rho} D_N)$. The definition of the 
properly subtracted operators necessary for the $\Delta I =1/2$ rule and for
$\epsilon'/\epsilon$ will be given  
in sections \ref{sec:DI=1/2} and \ref{sec:ep/e}.

\section{Penguin diagrams in perturbation theory}
\label{sec:penguins}
\begin{figure}[htb]
\begin{center}
\caption{\it{Penguin diagrams for four-fermion operators.}}
\label{fig:penguins}
\vskip 0.3cm
\includegraphics[height=5cm,width=10cm]{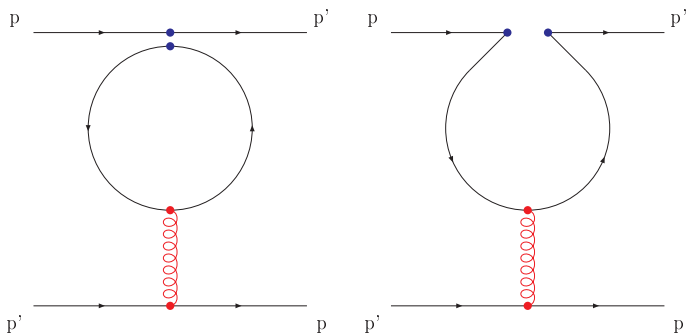}
\end{center}
\end{figure}

\begin{figure}[htb]
\begin{center}
\caption{\it{Feynman graphs for the vertex corrections to four-fermion operators.}}
\label{fig:gluonexch}
\vskip 0.3cm
\includegraphics[height=8cm,width=12cm]{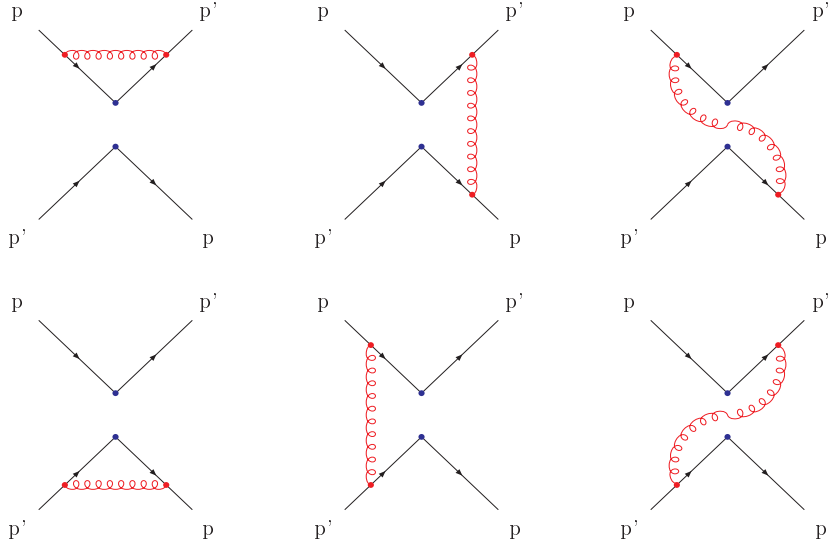}
\end{center}
\end{figure}

In this section we discuss the calculation of the penguin 
diagrams at one loop in perturbation theory.
The four-point Green's functions of a four-fermion operator 
between quark states are defined as 
\begin{equation}
G_i(x_1,x_2,x_3,x_4)=
\<q_1(x_1)\bar q_2(x_2) Q_i(0)
q_3(x_3)\bar q_4(x_4)\>\, ,  
\label{eq:G_Gamma(x)}
\end{equation}
where $x_1,x_3$ and $x_2,x_4$  are the coordinates of the  
outgoing and incoming quarks respectively.
Note that $G_i$ depends implicitly on the color
and Dirac indices carried by the external fermion fields. 
The Fourier transform of the Green's functions 
(\ref{eq:G_Gamma(x)}), with external momenta $p$  and $p'$ 
chosen as in Figs. \ref{fig:penguins} and \ref{fig:gluonexch}, 
is  defined as
\begin{equation}\label{eq:G_Gamma(p)}
G_i(p,p')
= \int dx_1 dx_2 dx_3 dx_4 
e^{-ip(x_1 - x_4)} e^{-ip'(x_3 - x_2)} G_{i}
(x_1,x_2,x_3,x_4)\; .
\end{equation}
The corresponding amputated correlation functions are defined as 
\be\label{eq:amputate}
\Lambda_i(p,p') = S^{-1}(p) S^{-1}(p') G_i(p,p')  S^{-1}(p') S^{-1}(p)\; ,
\ee
where $S(p)$ denotes the quark propagator in momentum space.
At one loop, the only diagrams which contribute to the mixing 
among dimension-six operators are 
the penguin diagrams in Fig.~\ref{fig:penguins}
and the vertex diagrams in Fig.~\ref{fig:gluonexch}. 
As for the $\Delta F =2$ operators in Ref.~\cite{capgiu},
it can be proven that the operators
in Eqs.~(\ref{eq:basis_lattice}) renormalize as
the corresponding local ones. Therefore
in the following all one-loop computations are performed
with local operators.

The vertex diagrams in Fig.~\ref{fig:gluonexch} have been
computed in Ref.~\cite{capgiu};  they are sufficient
for the renormalization of  
the $\Delta F = 2$ operators and do not give rise to 
mixing with lower-dimensional operators. The penguin diagrams in  
Fig.~\ref{fig:penguins} are computed here for the first time and 
constitute one of the main results of this paper.
The calculations involve the gauge-invariant quantity 
\be
{\cal P}_\mu (ak,am) = \int_{-\pi/a}^{\pi/a} \frac{d^4 l}{(2\pi)^4}
S^{(0)}_N(l-\frac{k}{2})
V^{(1)}_\mu(l+\frac{k}{2},l-\frac{k}{2})S^{(0)}_N(l+\frac{k}{2})\; , \qquad k
=p'- p \; ,
\ee
which does not depend on the flavor, spin and color
structures of the operators, and is the result of the internal quark loop 
once these structures have been factored out.
The tree-level overlap propagator $S^{(0)}_N(p)$ and the
quark-gluon vertex $V^{(1)}_\mu(p,p')$ are defined in Ref.~\cite{capgiu}.  
Eventually we will impose the $\RI$ renormalization scheme
and therefore we can perform all the computations 
in the massless limit. ${\cal P}_\mu (ak,0)$ has to be computed by 
expanding vertices and propagators to second order in $ak$,
since in the diagrams in Fig.~\ref{fig:penguins} 
it is contracted  with the gluon propagator $G(ak) \sim (ak)^{-2}$. 

For the analytic computations we have used FORM codes which we have 
developed specifically for this problem; the output has been 
integrated on a $60^4$ grid. Among other checks, we have computed 
${\cal P}_\mu (ak,0)$ using different assignments for 
the loop momenta. We also checked that the spectator quark line 
contributes only with its continuum expression.
The result for overlap fermions is 
\begin{equation}\label{eq:peng}
{\cal P}_\mu (ak,0) = i \frac{g_0}{16\pi^2} \Bigg[  
-\frac{1}{3} \log a^2 k^2 + B_{peng} \Bigg] \Big(\gamma_\mu k^2  - k \!\!\! / k_\mu  
\Big) ,
\end{equation}
where $B_{peng}$ is given in Table~\ref{tab-overlap-B} for several values of $\rho$. 
Notice the structure in $k$ which comes from gauge invariance. 
\begin{table}
\begin{center}
\begin{tabular}{||l|rrr|r||}
\hline\hline
$\rho$ &$B_\psi$ & $B_S$  & $B_V$ & $B_{peng}$ \\
\hline
                     0.2 &-235.80762 & 1.31942 & 1.52122 & 0.64996\\ 
                     0.3 &-150.61868 & 1.89625 & 1.52277 & 0.49361\\ 
                     0.4 &-108.19798 & 2.38060 & 1.52448 & 0.45000\\ 
                     0.5 & -82.86081 & 2.80522 & 1.52637 & 0.44558\\ 
                     0.6 & -66.05227 & 3.18782 & 1.52845 & 0.45796\\ 
                     0.7 & -54.10921 & 3.53927 & 1.53074 & 0.47845\\ 
                     0.8 & -45.20179 & 3.86686 & 1.53329 & 0.50317\\ 
                     0.9 & -38.31447 & 4.17577 & 1.53611 & 0.53024\\ 
                     1.0 & -32.83862 & 4.46989 & 1.53924 & 0.55875\\ 
                     1.1 & -28.38734 & 4.75224 & 1.54274 & 0.58826\\ 
                     1.2 & -24.70304 & 5.02527 & 1.54665 & 0.61863\\ 
                     1.3 & -21.60760 & 5.29104 & 1.55105 & 0.64992\\ 
                     1.4 & -18.97397 & 5.55135 & 1.55601 & 0.68235\\ 
                     1.5 & -16.70910 & 5.80783 & 1.56163 & 0.71635\\ 
                     1.6 & -14.74330 & 6.06201 & 1.56804 & 0.75261\\ 
                     1.7 & -13.02336 & 6.31544 & 1.57541 & 0.79229\\ 
                     1.8 & -11.50798 & 6.56970 & 1.58394 & 0.83759\\ 
\hline\hline
\end{tabular}
\caption{Perturbative values of the finite parts of the 
amputated correlation functions in the Landau gauge for the 
self energy, the scalar density and the vector current.  
The gauge-invariant penguin constant $B_{peng}$ defined in 
Eq.~(\ref{eq:peng}) is reported in the last column.}
\label{tab-overlap-B}
\end{center}
\end{table}
The $k \!\!\! /$ term vanishes, up to $O(g_0^2)$, if the equations
of motion are used for the spectator quarks \cite{bsd}, and the $k^2$ in the remaining structure 
$\gamma_\mu k^2$ disappears when combined with the gluon propagator.

We have used our codes to compute the penguin diagrams also in the Wilson case, 
\begin{eqnarray}
{\cal P}^{W}_\mu (ak,0) &=& i \frac{g_0}{16\pi^2} \Bigg[
-\frac{1}{3} \log a^2 k^2 + 0.5709352 \Bigg] 
\Big(\gamma_\mu k^2 - k \!\!\! / k_\mu  \Big)\\
&=&i \frac{g_0}{16\pi^2} \Bigg[  
-\frac{1}{3} \log \frac{a^2 k^2}{\pi^2} -0.1922181 \Bigg] 
\Big(\gamma_\mu k^2 - k \!\!\! / k_\mu  \Big) , \nonumber
\end{eqnarray}
and we are in agreement with the results by Bernard et al.~\cite{bsd}. 
It is interesting to note that the overlap result for $\rho=1$ is not too 
far from the Wilson number. In the Wilson case the lack of chiral 
symmetry causes, even in the massless case, the presence in 
${\cal P}^W_\mu (ak,0)$ of an additional power-divergent term  which contains 
tensor operators:
\begin{equation}\label{eq:sfigato}
-0.39847257 \, \frac{1}{a} \frac{g_0}{16\pi^2} \sigma_{\mu\lambda} k_\lambda .
\end{equation}
We have also observed that, as expected, for overlap fermions the coefficient of the
tensor term is zero (within our integration precision). This constitutes another test of the 
numerical accuracy of our integrals.

A generic local four-fermion operator can give rise 
to one or more of the following penguin types\footnote{We use the notation
$\bar{s} \Gamma {\cal T} d = \bar{s}^\alpha_i \Gamma_{ij} {\cal T}_{\alpha\beta} d^\beta_j $, 
$\Gamma$ and $\Gamma'$ are general Dirac matrices, ${\cal T}$ and 
${\cal T}'$ are color matrices ($\id$ or $t^A$), and $\alpha$, $\beta$ are color indices.}:
\ba
P_{A1} & = & (\bar{s} \Gamma {\cal T} q) 
            (\bar{q} \Gamma' {\cal T}' d) \nonumber\\
P_{A2} & = & (\bar{s}^\alpha \Gamma q^\beta) 
            (\bar{q}^\beta \Gamma' d^\alpha) \nonumber\\
P_{B1} & = & (\bar{s} \Gamma {\cal T} d )
            (\bar{q} \Gamma' {\cal T}' q) \\
P_{B2} & = & (\bar{s}^\alpha \Gamma d^\beta ) 
            (\bar{q}^\beta \Gamma' q^\alpha) \nonumber.
\ea
The Wick contractions which form the quark loop are always meant 
to be only between $q$ and $\bar{q}$, and at one loop they give 
\begin{equation}
\label{eq:4-penguins}
\begin{array}{ccccc}
{\cal I}_{A1} = -i g_0 G(ak) (
        \bar{s} \Gamma & {\cal P}_\mu (ak,am) \Gamma'  
         & {\cal T} t^A {\cal T'}
         & d)( \bar{f} t^A \gamma_\mu f) & \\
{\cal I}_{A2} = -i g_0 G(ak) (
       \bar{s} \Gamma & {\cal P}_\mu (ak,am)\Gamma'  
         &  \mbox{tr} (t^A)     
         & d)( \bar{f} t^A \gamma_\mu f) & =0 \\
{\cal I}_{B1} = -i g_0 G(ak) (
       \bar{s} \Gamma & \mbox{tr} (-{\cal P}_\mu (ak,am)\Gamma' ) 
         & {\cal T} \mbox{tr} (t^A {\cal T'}) 
         & d)( \bar{f} t^A \gamma_\mu f) & \\
{\cal I}_{B2} = -i g_0 G(ak) (
        \bar{s} \Gamma & \mbox{tr} (-{\cal P}_\mu (ak,am)\Gamma' ) 
         & t^A     
         & d)( \bar{f} t^A \gamma_\mu f) & ,
\end{array}
\end{equation}
where $G(ak)$ is the gluon propagator given in Ref.~\cite{capgiu} and 
$f$ is the spectator quark.
For the contractions $P_{A1}$ and $P_{A2}$ the relevant diagram is the one on the
right of Fig.~\ref{fig:penguins}, while for $P_{B1}$ and $P_{B2}$ it is the 
diagram on the left. 
Weak operators in which $q=d$ or $q=s$, since they have three quarks of the same
flavor, can undergo two Wick contractions instead of only one, therefore 
contributing with both diagrams to their renormalization factors. 

The 1-loop penguin corresponding to ${\cal I}_{A2}$ is zero because of the tracelessness of 
the $t^A$ matrices; when ${\cal T} = {\cal T}' = \id$ the penguin contribution
${\cal I}_{B1}$ also vanishes for the same reason.
Notice also that, since ${\cal P}_\mu$ enters in the expressions~(\ref{eq:4-penguins}) 
either as $\Gamma {\cal P}_\mu \Gamma'$ or as $\Gamma \mbox{tr} (-{\cal P}_\mu
\Gamma' )$, the contribution of the $\sigma_{\mu\lambda}$ term of
Eq.~(\ref{eq:sfigato}) in the Wilson
case vanishes for all operators considered in this paper.

\section{Renormalization pattern for the $\Delta I=1/2$ rule}
\label{sec:DI=1/2} 
For renormalization scales above the charm mass the bulk 
of the CP-conserving $\Delta I =1/2$ amplitudes 
comes from the operators
\be
{\cal O}_{\pm} = ({\cal Q}_1 - {\cal Q}^c_1) \pm ({\cal Q}_2 - {\cal Q}^c_2)\; .
\ee
The contributions which arise when the top quark is integrated out
are suppressed by a factor $\lambda_t/\lambda_u$ and can be 
neglected\footnote{The case with a charm integrated out can 
be easily obtained from the discussion in the next section.}.

Chiral symmetry forbids mixings with other dimension-six operators 
and the flavor structure forbids mixings between 
${\cal O}_{+}$ and ${\cal O}_{-}$. When  
chiral symmetry is preserved, the GIM mechanism ensures that 
the mixing with ${\cal Q}_\sigma$ and ${\cal Q}_m$ 
(Eqs.~(\ref{eq:Qsigma}) and (\ref{eq:Qm})) vanishes as 
$m_c^2-m_u^2$ when $m_c\rightarrow m_u$.
This can be shown as in the continuum: the penguin 
contractions of left-left four-fermion operators
take non-vanishing contributions only when 
the helicity, which is preserved through the massless quark lines, 
is flipped by an even number of mass insertions. 
Therefore in the 
overlap regularization the renormalized operators read as 
\ba\label{eq:rops} 
\widehat O_{\pm}(\mu) & = & Z_{\pm}(\mu a, g^2_0) \widetilde {\cal
O}_{\pm}(a) + O(a^2)\\
\widetilde {\cal O}_{\pm}(a) & = & {\cal O}_{\pm}(a) + (m^2_c -m^2_u) 
C_{\pm}^m(g_0^2) {\cal Q}_m(a)\nonumber \; ,
\ea
where $Z_{\pm}(\mu a, g^2_0)$ are $O(1)$ logarithmic divergent 
renormalization constants, $C_{\pm}^m(g_0^2)$ are finite mixing 
coefficients \cite{massimo} which are suppressed by a factor  $\alpha_s$
and the contributions proportional to ${\cal Q}_\sigma$ give
$O(a^2)$ effects.
$Z_{\pm}$ and $C_{\pm}^m$ can be determined by imposing 
two renormalization conditions which, in perturbation theory, 
would specify their two- and four-quark amputated Green's functions.

For $Z_{\pm}$ we opt for the $\RI$ renormalization scheme proposed 
in \cite{NP,RI}.  $Z^{\ri}_{\pm}$ are defined by imposing 
that the renormalized amputated Green's functions at large fixed  
Euclidean scale $\mu^2$, in the 
Landau gauge and in the massless limit, are equal to their 
tree-level values\footnote{In the Landau gauge this is equivalent to the 
use of the projectors as in Ref.~\cite{capgiu}.}, i.e.
\be\label{eq:ridef}
 Z^{\ri}_{\pm}(\mu a, g^2_0) 
\Lambda_{\pm}(p,p')\Big|_{p^2={p^{'2}}=\mu^2} = \Lambda_\pm^{(0)}(p,p')
\Big|_{p^2={p^{'2}}=\mu^2} \; .
\ee 
At one loop the relevant contributions to $Z^{\ri}_{\pm}$ 
come from the vertex diagrams\footnote{Throughout this paper
the $\Lambda_i(p,p')$'s for the four-fermion operators include only diagrams
which contribute to the mixing among dimension-six operators.} in
Fig.~\ref{fig:gluonexch}, while the penguin diagrams cancel 
in the massless limit due to the GIM mechanism. 
Using the results in Ref.~\cite{capgiu} we obtain
\be
Z_{\pm}^{\ri} = 1-\frac{g_0^2}{16 \pi^2}
\left[\frac{\gamma^{(0)}_{\pm}}{2} \log(\mu^2 a^2) + B_{\pm} + 2 
\frac{N_c^2-1}{2N_c} B_\psi\right]\; ,
\ee
where 
\ba
\gamma^{(0)}_{+}  = 4  & \qquad &   B_{+} = \frac{10}{3}B_V -\frac{2}{3} B_S\nonumber\\
\gamma^{(0)}_{-}  = -8  & \qquad &  B_{-} = \frac{4}{3}(B_S + B_V)\; . 
\ea 
The constants $B_\psi$, $B_S$ and $B_V$, which enter the renormalization
of the quark propagator, the scalar density and the vector current, were computed 
in \cite{vicari,capgiu} for several values of $\rho$ and are reported 
in Table~\ref{tab-overlap-B} for completeness. The anomalous dimensions 
$\gamma^{(0)}_{\pm}$ are in agreement with those in Refs.~\cite{buras,reina} and
the matching coefficients between $\RI$ and one of the 
$\MSbar$ schemes can be found in \cite{RI}. 

The mixing coefficients $C^{(\pm)}_m$ can be determined by
imposing suitable renormalization conditions on two-quark correlation 
functions computed at two loops in perturbation theory. 
As in the continuum,  these coefficients are not needed for the physical
$K\rightarrow \pi\pi$ matrix elements. For
$K\rightarrow \pi$ amplitudes we will show that they can be fixed 
non-perturbatively by imposing suitable renormalization conditions 
on the $K\rightarrow 0$ matrix elements.

The $K\rightarrow \pi\pi$ matrix elements
\ba\label{eq:matrelem} 
\langle \pi\pi| \widehat O_{\pm}(\mu) |K\rangle 
 =  Z_{\pm}(\mu a, g^2_0) \langle \pi\pi| \left[{\cal O}_{\pm}(a) + (m^2_c -m^2_u) 
C_{\pm}^m(g_0^2) {\cal Q}_m(a)\right]|K\rangle \; 
\ea
can be extracted directly from the four-point functions once the 
final state interactions have been properly taken 
into account \cite{LaurentLuscher}-\cite{testa_new}. 
Following \cite{testa}, the on-shell matrix elements of 
the pseudoscalar density ${\cal P} = \bar s \bar \gamma_5 d$ satisfies 
the Axial Ward Identity
\be
Z_A \langle \pi\pi|\partial_\mu {\cal A}_{\mu}|K\rangle = 
(m_s + m_d) \langle \pi\pi|{\cal P}|K\rangle + O (a^2)\; ,
\ee
where ${\cal A}_\mu = \bar s \gamma_\mu\bar \gamma_5 d$ and 
$Z_A$ is its renormalization constant, and up to $O(a^2)$ effects 
$ \langle \pi\pi|{\cal P}|K\rangle$ vanishes when the momentum $\Delta p $ inserted by the 
operator is zero.
Eventually the physical matrix elements 
in the $\RI$ scheme read 
\be
\langle \pi\pi| \widehat O_{\pm}(\mu) |K\rangle =  
Z^{\ri}_{\pm}(\mu a, g^2_0) \langle \pi\pi| {\cal
O}_{\pm}(a)|K\rangle + O(a^2)\; .
\ee
Even if the momentum inserted is not zero,
the contribution proportional to $\Delta p$ is 
finite and suppressed by a factor $\alpha_s$.
Therefore the extrapolation to the physical 
point $\Delta p =0$ would not be a problem.
The direct computation described above does not
rely on chiral perturbation theory
and in principle the use of $O(a)$ improved operators renders 
the computation of the matrix elements sensitive to higher-order 
chiral contributions. The two main
advantages of the overlap regularization with respect to the Wilson fermions
are that the mixings with dimension-three operators are finite even 
at non-zero momentum transfer ($\Delta p \neq 0$) and 
the improvement of the bilinear and four-fermion operators 
can be performed without tuning any parameters. 
The main difficulties of this 
method are that the final state interactions have to be taken properly 
into account and the computations of the four-point 
functions are technically quite complicated which could lead 
to noisy signals. 

These difficulties can be avoided if one 
applies the method proposed 
in \cite{bs,old_lat}, where 
it has been suggested  to use 
chiral perturbation theory to relate $K\rightarrow \pi\pi$ 
to $K\rightarrow \pi$ matrix elements and to compute the latter
on the lattice. At the leading order in $\chi$PT\footnote{For higher
order results see Ref.\cite{golty}.} 
\ba
\langle 0| {\cal O}_{\pm} | K^0\rangle & = & i \delta_{\pm} (m_K^2-m_\pi^2)
\label{eq:primachiPT}\\
\langle \pi^+ (p) | {\cal O}_{\pm} | K^+(q)\rangle & = & 
\alpha_{\pm} \; p \cdot q \; 
-\delta_{\pm}\frac{m_K^2}{f_\pi}\; ,\label{eq:secochiPT}\\
\langle \pi^+ \pi^- | {\cal O}_{\pm} | K^0\rangle & = &  
i \alpha_{\pm} \frac{m_K^2-m_\pi^2}{f_\pi}\; ,\label{eq:lastchiPT}
\ea
and $\alpha_{\pm}$ can be computed by extracting   
$\langle \pi^+ (p) | {\cal O}_{\pm} | K^+(q)\rangle$ 
from the three-point functions and 
fitting the results  as a function of $p \cdot q$ and $m_K^2$.
The unphysical contribution proportional to $\delta_{\pm}$
is not divergent. This is one of the main results of this paper. 
$\delta_{\pm}$ can also be subtracted by fixing  
$C_{\pm}^m$ in such a way that \cite{bdspw}
\be
\langle 0 | \widetilde O_{\pm} | K^0\rangle = 0\; ,
\ee
i.e.
\be
C_{\pm}^m =  -\frac{\delta_{\pm}}{\delta_p
  (m_c^2-m_u^2)}\frac{m_K^2-m_\pi^2}{m_s-m_d}
\ee
with 
\be
\langle 0 | {\cal P} | K^0\rangle = i\delta_p\; .
\ee
The comparison of the two methods is a check of the whole procedure.
Since only single-particle states are involved there 
are no problems with final state interactions and 
the three-point functions are technically easier to simulate. 

From the considerations reported above, we can appreciate the advantages of 
using Neuberger's fermions to compute the matrix elements necessary 
for the $\Delta I=1/2$ rule:
\begin{itemize}
\item the GIM mechanism combined with chiral 
      symmetry constrains the mixing coefficients with 
      lower-dimensional operators to be proportional to $m_c^2-m_u^2$; 
\item the mixing coefficients with parity-conserving 
      and parity-violating components of lower-dimensional operators 
      are the same and are multiplied by $m_s - m_d$ and $m_s + m_d$ respectively;
\item ${\cal O}^{\pm}(a)$ do not mix with dimension-six operators with different 
      chiralities;
\item the $O(a)$ improvement of the operators does not require
      any tuning of parameters.     
\end{itemize}
On the contrary, in the Wilson regularization the 
parity-conserving components of ${\cal O}^{\pm}$ 
can mix with dimension-six operators of different chiralities, 
their mixing with lower-dimensional operators is no longer 
proportional to $m_s + m_d$ and the GIM mechanism gives only a factor
$m_c-m_u$. Therefore the contributions of the chromomagnetic 
operator have to be subtracted and the scalar operator mixes with
quadratically divergent coefficients. Finally for Wilson fermions 
the $O(a)$ improvement of the four-fermion composite operators 
requires tuning many parameters.

\section{Renormalization pattern for $\epsilon'/\epsilon$}
\label{sec:ep/e}
For CP violating processes in kaon decays, when the top quark is integrated out, the 
GIM mechanism is not operative and the presence
or not of an active charm does not substantially modify  
the mixing pattern of the relevant operators. Therefore, 
in this section we will consider the case with the charm 
quark integrated 
out\footnote{The case with a dynamical 
charm can be easily obtained from the results reported in this 
and in the previous section.}. The corresponding $\Delta S =1$
effective Hamiltonian can be obtained from Eq.~(\ref{eq:eh})
and is reported, for example, in Ref.~\cite{buras}. 

Analogously to the continuum, the operator 
\be
{\cal Q}_{27} = 2 {\cal Q}_2 + 3 {\cal Q}_1 - {\cal Q}_3
\ee
does not mix with lower-dimensional operators and 
it is multiplicatively renormalizable: 
\be
\widehat Q_{27}(\mu) = Z_{27}(\mu a, g^2_0){\cal Q}_{27}(a)\; .
\ee
At one loop in perturbation theory only the diagrams in 
Fig.~\ref{fig:gluonexch} give contributions to 
$Z_{27}(\mu a, g^2_0)$ and in the $\RI$ scheme, defined analogously 
to Eq.~(\ref{eq:ridef}), we obtain   
\be
Z^{\ri}_{27}(\mu a) =  1-\frac{g_0^2}{16 \pi^2}
\left[\frac{\gamma^{(0)}_{27}}{2} \log(\mu^2 a^2) + B_{27} + 2 
\frac{N_c^2-1}{2N_c} B_\psi\right]\; ,
\ee
with
\ba
\gamma^{(0)}_{27} = 4 \qquad B_{27} =  \frac{10}{3} B_V -\frac{2}{3}B_S \; .
\ea
Also in this case chiral perturbation theory can be used 
to relate $K\rightarrow \pi\pi$ to $K\rightarrow \pi$ matrix elements and 
the latter can be extracted from the three-point functions 
computed numerically.

The mixing pattern of the QCD penguin operators 
\ba
{\cal Q}_{3,5} & = & ({\bar s}\bar \gamma^\mu_L d)
\sum_q ({\bar q}\bar \gamma^\mu_{L,R} q)\\
{\cal Q}_{4,6} & = & ({\bar s}^{\alpha}\bar \gamma^\mu_L d^{\beta})
\sum_q ({\bar q}^{\beta}\bar \gamma^\mu_{L,R} q^{\alpha})\; \nonumber
\ea
is
\ba\label{eq:belle}
\widehat Q_i (\mu ) & = & \widehat Z_{ij} (\mu a,g_0^2) \widetilde{\cal Q}_j(a)\qquad
\qquad \qquad \qquad i,j=3,\dots,6\\
\widetilde{\cal Q}_j(a) & = & {\cal  Q}_j(a) +  
C_{j}^\sigma(g_0^2) {\cal Q}_\sigma(a) + C_{j}^m(g_0^2) {\cal Q}_m(a)\nonumber \; .
\ea
At one loop the overall renormalization constants $Z_{ij} (\mu a,g_0^2)$ 
take contributions from the penguin  and vertex diagrams in 
Figs.~\ref{fig:penguins} and \ref{fig:gluonexch}. Imposing the 
$\RI$ renormalization conditions 
\be
 Z_{ij}^{\ri} (\mu a,g_0^2) \Lambda_{j}(p,p')\left|_{p^2 = p^{'2}=\mu^2}\right.  =
\Lambda_i^{(0)}(p,p') \left|_{p^2 = p^{'2}=\mu^2} \right.
\ee
we obtain
\be\label{eq:ultima_fatica}
Z_{ij}^{\ri} (\mu a,g_0^2) = 1-\frac{g_0^2}{16 \pi^2}
\left[\frac{\gamma^{(0)}_{ij}}{2} \log(\mu^2 a^2) + B_{ij} + 2\delta_{ij} 
\frac{N_c^2-1}{2N_c} B_\psi\right]\; 
\ee
with $i,j=3,\dots 6$, 
\be\label{eq:dimanom}
\widehat \gamma^{(0)} = \pmatrix{ 
\displaystyle -\frac{22}{9}  & \displaystyle \frac{22}{3}     &
\displaystyle -\frac{4}{9}  &\displaystyle \frac{4}{3}     \cr
& & & \cr
\displaystyle 6-\frac{2f}{9} &\displaystyle -2+\frac{2f}{3}  &
\displaystyle -\frac{2f}{9} &\displaystyle 
\frac{2f}{3}    \cr
& & & \cr
     0         &      0          &      2       &     -6         \cr
& & & \cr
\displaystyle -\frac{2f}{9}  &\displaystyle \frac{2f}{3}     &
\displaystyle -\frac{2f}{9} &\displaystyle 
-16+\frac{2f}{3}\cr}\; , 
\ee

\be\label{eq:bbiii}
\widehat B = \pmatrix{ 
\displaystyle \frac{7}{3}B_V + \frac{B_S}{3}  + \frac{2}{3}B_{peng}&
\displaystyle B_V-B_S -2B_{peng}& \displaystyle
\frac{2}{3}B_{peng}&\displaystyle -2B_{peng}\cr
& & & \cr
\displaystyle B_V-B_S +\frac{f}{3}B_{peng}& \displaystyle \frac{7}{3}B_V + 
\frac{B_S}{3}  -f B_{peng}& \displaystyle
 \frac{f}{3}B_{peng}&\displaystyle  -f B_{peng}     \cr
& & & \cr
 0 &   0    & \displaystyle 3 B_V - \frac{B_S}{3}& -B_V + B_S \cr
& & & \cr
\displaystyle \frac{f}{3} B_{peng}  & \displaystyle -f B_{peng}      & 
\displaystyle \frac{f}{3} B_{peng}  &\displaystyle \frac{8}{3}B_S-f B_{peng}     \cr} \; , 
\ee
where\footnote{The case of a dynamical charm is obtained
with $f=4$.} $f=3$. Notice that all the mixing coefficients
in Eq.~(\ref{eq:ultima_fatica}) depend on four constants only.

The mixing coefficients $C_{j}^\sigma$ are finite, 
can be computed in perturbation theory and are relevant only if one is interested 
in higher-order 
contributions in $\chi$PT. We postpone the computation of $C_{j}^\sigma$
to a future publication.

The mixing coefficients $C_{j}^m$ are quadratically divergent in the lattice
spacing $a$ and have to be computed non-perturbatively. 
As in the previous section, they can be fixed by imposing the renormalization 
condition \cite{bdspw}
\be
\langle 0|\widetilde {\cal Q}_i | K^0\rangle = 0 \; .
\ee
The $K\rightarrow \pi\pi$ or, if $\chi$PT is used, the 
$K\rightarrow \pi$ matrix elements can be computed analogously to  
section \ref{sec:DI=1/2}. The $K\rightarrow \pi$ matrix elements require 
one power-divergent subtraction only and there are no extra mixings 
with dimension-six operators of different chirality.
For Wilson fermions, on the contrary, 
mixings with dimension-six operators of different chiral behavior are
allowed and two power-divergent subtractions are required.

\section{Conclusions}\label{sec:conclusions} 
We have analyzed various possibilities for computing the matrix elements 
relevant for the $\Delta I =1/2$ rule and for $\epsilon'/\epsilon$
with Neuberger's fermions. When chiral symmetry is preserved 
at finite lattice spacing the renormalization factors of the 
parity-violating and parity-conserving components of the four-fermion 
operators are the same. Therefore, contrary to the Wilson 
case, there are no extra difficulties in extracting the matrix elements
of the properly renormalized parity-conserving operators.
We have shown that the $K\rightarrow \pi\pi$ and, 
if chiral perturbation theory is used, 
$K\rightarrow \pi$ matrix elements 
for the  $\Delta I =1/2$ rule can be computed 
without any power subtractions. This is a consequence of the
exact chiral symmetry at finite lattice spacing and of the 
GIM mechanism which is quadratic in the 
quark masses. The analogous
matrix elements for $\epsilon'/\epsilon$ require 
only a power subtraction which can be performed non-perturbatively
using on-shell $K\rightarrow 0$ matrix elements.
All the required renormalization coefficients among dimension-six operators
have been computed at one loop in perturbation theory and depend on four
numerical constants only.
We believe that the overlap-Dirac operator is a promising regularization
to solve these long-standing problems.

\section*{Acknowledgments}
L.~G.~warmly thanks G.~Martinelli, C.~Rebbi and M.~Testa for many
illuminating discussions and S.~Sharpe for an interesting conversation. 
S.~C.~has been supported in part by the U.S. Department of Energy (DOE) under cooperative 
research agreement DE-FC02-94ER40818. L. G. has been supported in part under 
DOE grant DE-FG02-91ER40676.


\begin{thebibliography}{99}
\bibitem{epp/ep_exp} 
A. Alavi-Harati et al.,
Phys.~Rev.~Lett. 83 (1999) 22;\\
V. Fanti et al.,
Phys.~Lett.~B465~(1999)~335;\\
A. Ceccucci, CERN Particle Physics Seminar (29 February 2000). 
\bibitem{NA31_old}
H.~Burkhardt et al.,
Phys. Lett. B206 (1988) 169;\\
G.~D.~Barr et al.,
Phys. Lett. B317 (1993) 233.
\bibitem{luscher2}
M.~L\"uscher,
Nucl. Phys. B549 (1999) 295.
\bibitem{golterman}
M.~Golterman,
Plenary talk given at Lattice 2000, Bangalore - India, August 2000,\\ 
hep-lat/0011027, and references therein.
\bibitem{deltaI=1/2}
M.~K. Gaillard, B.~W. Lee, 
Phys.~Rev.~Lett. 33 (1974) 108;\\
G.~Altarelli, L.~Maiani,
Phys.~Lett.~B52~(1974)~351.
\bibitem{buras}
A. J. Buras, M. Jamin, M. E. Lautenbacher, P. H. Weisz,
Nucl. Phys. B370 (1992) 69;\\
Nucl. Phys. B400 (1993) 37;\\ 
A. J. Buras, M. Jamin, M. E. Lautenbacher, 
Nucl. Phys. B400 (1993) 75. 
\bibitem{reina}
M. Ciuchini, E. Franco, G. Martinelli, L. Reina, 
Nucl. Phys. B415 (1994) 403. 
\bibitem{silvester}
S.~Bosch et al.,
Nucl. Phys. B565 (2000) 3.
\bibitem{ep/e}
M.~Ciuchini, E.~Franco, L.~Giusti, V.~Lubicz, G.~Martinelli,
Nucl. Phys. B573 (2000) 201.  
\bibitem{pp}
E.~Pallante, A.~Pich,
Phys. Rev. Lett. 84 (2000) 2568 and Nucl. Phys. B592 (2000) 294;\\
A.~J.~Buras et al., 
Phys. Lett. B480 (2000) 80.
\bibitem{boc}
M.~Bochicchio et al., 
Nucl. Phys. B262 (1985) 331.
\bibitem{shape1} S.~Sharpe et al., 
Nucl.~Phys. B286 (1987) 253;\\
S. R.~Sharpe, A.~Patel, 
Nucl.~Phys. B417 (1994) 307.
\bibitem{tassos}
A. Donini, V. Gim\'enez, G. Martinelli, M. Talevi, A. Vladikas, 
Eur.~Phys.~J.~C10 (1999)~121. 
\bibitem{sharpe}
T.~Bhattacharya, R.~Gupta, S. R.~Sharpe, 
Phys.~Rev. D55 (1997) 4036. 
\bibitem{japanbk_ks}
JLQCD Collaboration (S. Aoki et al.),
Phys.~Rev.~Lett. 80 (1998)~5271. 
\bibitem{japanbk}
JLQCD Collaboration (S. Aoki et al.), 
Phys.~Rev.~D60 (1999)~034511. 
\bibitem{NoiDELTAS=2}
L. Conti et al.,  
Phys.~Lett.~B421~(1998)~273.\\
C.~R.~Allton et al., 
Phys. Lett. B453 (1999)~30.\\
A. Donini, V. Gim\'enez, L. Giusti, G. Martinelli,
Phys. Lett. B470 (1999)~233.  
\bibitem{CKM}
F.~Caravaglios, F.~Parodi, P.~Roudeau, A.~Stocchi,
hep-ph/0002171.
\bibitem{CKM_beyond}
L.~Giusti, A.~Romanino, A.~Strumia,
Nucl.~Phys.~B550~(1999)~3; \\ 
R.~Barbieri, L. J. Hall, A. Romanino,
Nucl.~Phys.~B551 (1999) 93. 
\bibitem{bs}
C.~Bernard et al., 
Nucl.~Phys.~B (Proc. Suppl.)~4~(1988)~483.
\bibitem{old_lat}
L.~Maiani, G.~Martinelli, G.~C.~Rossi, M.~Testa, 
Nucl.~Phys.  B289 (1987) 505.
\bibitem{blum}
T. Blum,  
Talk given at Lattice 2000, Bangalore - India, August 2000,
hep-lat/0011042;\\
CP-PACS Collaboration (A. Ali Khan et al.),
Talk given at Lattice 2000, Bangalore - India, August 2000, 
hep-lat/0011007. 
\bibitem{maiani_testa} 
L.~Maiani, M.~Testa,
Phys. Lett. B245 (1990) 585. 
\bibitem{LaurentLuscher} 
L.~Lellouch, M.~L\"uscher,
hep-lat/0003023. 
\bibitem{sacmarti}
M.~Ciuchini, E.~Franco, G.~Martinelli, L.~Silvestrini,
Phys. Lett. B380 (1996) 353. 
\bibitem{testa_new}
M. Testa, 
Talk presented at ICHEP 2000, Osaka - Japan, hep-lat/0010020. 
\bibitem{neub1}
H.~Neuberger,
Phys. Lett. B417 (1998)~141;\\
H.~Neuberger,
Phys. Lett. B427 (1998)~353.
\bibitem{hasenfratz2}
P. Hasenfratz,
Nucl. Phys. B525 (1998)~401. 
\bibitem{luscher}
M.~L\"uscher,
Phys. Lett. B428 (1998) 342.
\bibitem{GW}
P.~H.~Ginsparg, K.~G.~Wilson, 
Phys. Rev. D25 (1982) 2649.
\bibitem{neub0}
R.~Narayanan, H.~Neuberger,
Phys. Lett. B302 (1993)~62;\\
R.~Narayanan, H.~Neuberger, Nucl. Phys. B443 (1995)~305.
\bibitem{pilar}
P.~Hern\'andez, K.~Jansen, M.~L\"uscher, 
Nucl.~Phys. B552 (1999)~363. 
\bibitem{simulazioni}
H. Neuberger, 
Phys. Rev. Lett. 81 (1998) 4060;\\
R.~G.~Edwards, U.~M.~Heller, R.~Narayanan,
Nucl. Phys. B540 (1999) 457;\\
A.~Bode, U.~M.~Heller, R.~G.~Edwards, R.~Narayanan,
hep-lat/9912043;\\
P.~Hern\'andez, K.~Jansen, L.~Lellouch,
hep-lat/0001008;\\
R.~Narayanan, H.~Neuberger,
Phys. Rev. D62 (2000) 074504;\\
S. J. Dong, F. X. Lee, K. F. Liu, J. B. Zhang, 
Phys.~Rev.~Lett. 85 (2000)~5051;\\ 
L. Giusti, C. Hoelbling, C. Rebbi, 
Nucl. Phys. (Proc. Suppl.) 83-84 (2000)~896 and \\
hep-lat/0011014;\\ 
T.~DeGrand,
Phys.~Rev.~D63~(2001)~034503. 
\bibitem{capgiu}
S. Capitani, L. Giusti,
Phys. Rev. D62 (2000) 114506. 
\bibitem{qcdsf}
S.~Capitani, M.~G\"ockeler, R.~Horsley, P.~E.~L.~Rakow, G.~Schierholz,\\ 
Phys.~Lett.~B468~(1999)~150; \\
S.~Capitani et al., 
Nucl. Phys. B593 (2001) 183.
\bibitem{aoki}
S.~Aoki, Y.~Kuramashi,
Phys.~Rev. D63 (2001)~054504. 
\bibitem{NP}
G. Martinelli, C. Pittori, C.T. Sachrajda, M. Testa, A. Vladikas, 
Nucl. Phys. B445 (1995)~81.
\bibitem{luscher_np}
M.~L\"uscher, S.~Sint, R.~Sommer, H.~Wittig,
Nucl. Phys. B491 (1997) 344. 
\bibitem{masmur}
A.~Masiero, H.~Murayama,
Phys. Rev. Lett. 83 (1999) 907.
\bibitem{narayanan}
R. Narayanan,
Phys. Rev. D58 (1998) 97501.
\bibitem{luscher3}
M.~L\"uscher,
JHEP  0006 (2000) 028. 
\bibitem{japan}
M.~Ishibashi, Y.~Kikukawa, T. Noguchi, A. Yamada, 
Nucl. Phys. B576 (2000) 501. 
\bibitem{vicari}
C. Alexandrou, E. Follana, H. Panagopoulos, E. Vicari,
Nucl. Phys. B580 (2000) 394. 
\bibitem{stefano_neub}
S.~Capitani,
Nucl. Phys. B592 (2001) 183 and Nucl. Phys. B597 (2001) 313.
\bibitem{bdspw}
C. Bernard, T. Draper, H. D. Politzer, A. Soni, M. B. Wise,
Phys. Rev. D32 (1985) 2343.
\bibitem{bsd}
C. Bernard, T. Draper,  A. Soni,
Phys. Rev. D36 (1987) 3224. 
\bibitem{massimo}
M.~Testa,
JHEP 9804 (1998) 002.
\bibitem{RI}
M. Ciuchini, E. Franco, G. Martinelli, L. Reina, L. Silvestrini,
Z. Phys. C68 (1995) 239.
\bibitem{testa}
C. Dawson et al.,
Nucl. Phys. B514~(1998)~313. 
\bibitem{golty}
M.~Golterman,
hep-ph/0011084;\\
M.~Golterman, E.~Pallante,
JHEP 0008 (2000) 023.
\end{thebibliography}
\end{document}